# Time-Symmetric Physics: A Radical Approach to the Decoherence Problem

Paul J. Werbos*
ECCS Division, National Science Foundation

**Abstract**

The most powerful form of quantum learning system possible would somehow learn the parameters W of a quantum system $\underline{f}(\underline{X}, W)$, for $\underline{f}$ representing the largest, most powerful set of possible input-output relations. This paper addresses the issue of how to enlarge the set represented by $\underline{f}$, by using a new formulation of time-symmetric physics to model analog quantum computers based on spin and by exploring possible sources of backwards-time free energy so as to address problems of decoherence and dissipation.

## 1 Introduction

Quantum learning machines, like all other forms of quantum computing and quantum communication, are limited by the long-standing problem of decoherence.

In essence, all types of computing may be thought of as families of functions, $\underline{Y}=\underline{f}(\underline{X}, W)$, where $\underline{f}$ represents the set of available functions, where $\underline{X}$ and $\underline{Y}$ are the vectors (or graphs) of inputs and outputs, and W is a member of the set of allowed parameter and programming choices. From the work of David Deutsch [1], we know that digital quantum computing, based on qubits built up from |0> and |1>, offers us a much larger set of available functions in principle than what we can get from standard universal Turing machines. From the work of Hava Siegelmann [2], we know that analog neural network computing offers us a larger set of functions (at least within a cost constraint) than standard digital Turing computing. Ultimately, in quantum learning, we want to find ways to learn W, so as to perform useful tasks, for functions $\underline{f}$ in the largest available set of possibilities – the set of analog quantum computing machines.

Before we can unleash the full power of quantum learning, we must first learn how to implement and model the full range of functions $\underline{f}$ available to us. This paper will discuss how new concepts in time-symmetric physics have a substantial impact on the choice of functions $\underline{f}$ available to us for quantum computing and quantum communication, in two ways. First, it affects the way we can model and design quantum information systems (QIS), when we use spin or polarization as an analog variable. Second, it offers a new way to try to reduce decoherence, a practical problem which seriously limits the choices of function $\underline{f}$ available in the real world, in all forms of QIS [3]. The first of these developments is far more mature than the second, and it provides an introduction to the physics underlying both approaches.

The actual development of learning algorithms and architectures to attach to these more powerful functional forms is a complicated issue, well beyond the scope of this paper. Development and understanding of the physical functional forms logically precedes our ability to train them.

## 2. Analog QIS and Time-Symmetric Physics

Quantum computing with photons is now one of the leading approaches to implementing QIS [3]. In traditional digital quantum computing, |0> and |1> are implemented as |H> and |V>, horizontal and vertical polarizations of a photon. This could be extended, of course, to QIS with other photon-like bosons, such as polaritons or plasmons or even excitons, which may allow a greater density of features on a chip, as this technology is perfected. In order to upgrade this to full analog QIS, the obvious approach is to compute with the full range of polarizations, |θ>, not just |H> and |V>.

Would this extension give us the same kind of growth in power that Siegelmann [2] has proven for classical systems? I cannot offer a proof here, but I can cite two very exciting new examples [4,5] from the field of quantum communication. Based on this work, it is now well established that new types of code, based on the full power of analog QIS, offer a substantial improvement in what can be achieved in unbreakable codes, beyond what is possible

---
[1] [1]The views herein represent no one's official views, but the paper was written on US government time

with traditional digital QIS. In general, it would be surprising if additional fundamental degrees of freedom did not offer fundamental new capabilities.

Just a few years ago, the QIS community was much less optimistic about optical quantum computing than it is now. For example, many argued that optical quantum computing would be limited to a kind of "linear quantum computing," which could only handle a subset of the capabilities envisioned by Deutsch, because of technology limitations at the time. The story changed when the Joint Quantum Institute at Maryland invented a new digital spin gate for photons, suitable for more general quantum logic.

What actually happens, however, when we build large networks of analog spin gates (polarizer/amplifiers?) for use in QIS, when there is a high degree of entanglement across many photons? Until we fully understand what happens, physically, with networks of polarizers involving three or four entangled photons, we will not be ready to design systems exploiting tens or hundreds of them. At some level, the task before us now in the mainstream is to really nail down that story, first; later sections of this paper could be seen as speculative until that is fully taken care of.

Despite all the excitement about systems involving many, many qubits, there are only three groups which I am aware of in the world which have actually, physically produced at least the basic general three-photon entangled states of Greenberger, Horne and Zeilinger [6]: (1) Zeilinger's group itself; (2) the group of Yanhua Shih at the University of Maryland Baltimore Campus; (3) the new group led by one of Zeilinger's former students, now in Sichuan province. At the recent SPIE conference on quantum information science and technology [7] and at Princeton workshop on quantum optics and noise [8], I asked whether anyone knew of any other groups; they had hopes that other groups might catch up, but did not know of any who have succeeded as yet.

The essential problem here is that there are two radically different ways to model these kinds of systems, which lead to very different predictions. One way is the traditional Copenhagen model, which assumes that the polarizer implements a "collapse of the wave function" [7]. Another way is based on a new stochastic path formulation of physics [9], which implements fundamental new concepts in time-symmetric physics [10]. The two different approaches do lead to the same predictions for the case of two entangled photons [7], as in the classic Bell's Theorem experiments illustrated in figure 1:

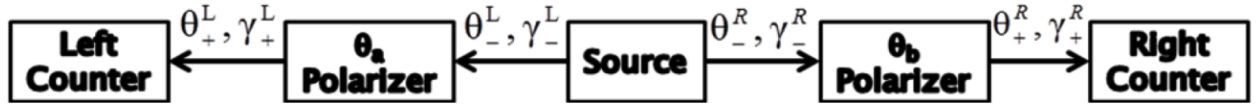

Figure 1. Core structure and notation for the first Bell's Theorem experiments

However, for the triphoton experiment illustrated in figure 2

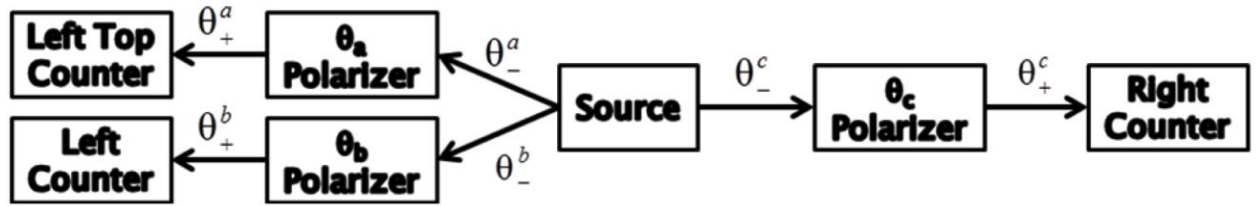

Figure 2. A possible triphoton experiment

where the source emits the GHZ state defined by

$$\mathcal{Y} = \frac{1}{\sqrt{2}}\left[|0\rangle_a|0\rangle_b\left|\frac{p}{2}\right\rangle_c + \left|\frac{p}{2}\right\rangle_a\left|\frac{p}{2}\right\rangle_b|0\rangle_c\right] \quad , \quad (1)$$

and where the outgoing photons reach the polarizers on the left before they reach the right, the Copenhagen method predicts [7] a relative three-photon counting rate of

$$R_3/R_0 = 1/2 \,(\cos\theta_a \cos\theta_b \sin\theta_c + \sin\theta_a \sin\theta_b \cos\theta_c)^2 \,. \quad (2)$$

but time-symmetric physics predicts:

$$R_3/R_0 = k \cos^2(\theta_c - \theta_a - \theta_b) \quad (3)$$

The first crucial experiment to discriminate between these two predictions is now underway at Maryland [11]. Until that experiment is completed, and replicated elsewhere, we do not yet know for sure which theory of physics is correct. However, the underlying logic of time-symmetric physics is so compelling, in my view [7-10], and the "collapse of the wave function" so artificial, that it makes some sense to look ahead, and ask what more could be derived from the new approach.

## 3. Decoherence, Time Symmetry and the Arrow of Time in Thermodynamics

Intuitively, if a quantum computer requires a time of $\tau_1$ to perform a simple operation, and if it can operate for a time of only $\tau_2$ because of decoherence effects, then it can only perform a program which completes in $\tau_2/\tau_1$ steps. When $\tau_2/\tau_1$ is not large enough, the effective power of quantum computers is very limited, with or without learning.

Mainstream research in quantum computing has developed two major techniques to address decoherence: (1) quantum error correction (QEC) and quantum nondemolition measurement (QND). These help, but the power of quantum computing is still limited by decoherence more than anything else. For decades, it has seemed that a new "breakthrough on decoherence" is announced every year, without changing this basic story.

It is interesting to ask whether the new models of measurement in section 2 might somehow be used to enhance QND. However, this section will focus on a more radical, complementary possibility.

### 3.1. Setting the Stage: Cavity QED

The problem of decoherence seems to be based on very fundamental physical limits, related to dissipation and noise. Yet many years ago, lasers also seemed to be facing a fundamental limit in trying to reduce noise, due to spontaneous emission, something which is unavoidable in the canonical, normal form version of quantum electrodynamics (KQED [12]). Practical experimenters argued that a laser in a cavity might not emit so many unwanted stray photons, because such photons would not have a good place to go after being emitted by the atom. The issue was hotly debated by theorists, but experimentalists prove that they were right; reduced noise lasers are now one of the most important tools available in practical engineering.

Theorists adapted to the new situation by developing a version of Cavity QED [13,14,15] which assumes that the vacuum state of our universe is fully of energy – zero point energy whose total density at any point in space is infinite (or $10^{117}$J/meter, strange enough). Some theorists might say: "So what if we introduce one more infinite quantity? We do that all the time with renormalization. Infinity minus infinity equals whatever we want it to equal." And in fact, this is consistent with the Feynman path version of QED (FQED[12]). Zero point energy is often cited as the basis of the Casimir effect – even though Landau showed long ago that the measured "Casimir effect" between two metal plates is precisely equal to the predictions from vanderwaals forces; the addition residual, predicted by zero point energy, is exactly zero, to the most accurate measurements available.

In [10], I propose an alternative explanation of this effect, which does not require the assumption of zero point energy. Very simply, a free atom emits a photon only when it "sees" a potential absorber available to absorb it in the future. This seems impossible in classical time-forwards intuition, but it fits very naturally with the kind of mathematics one uses to predict time-symmetric systems [7,9]. It also fits with examples like nuclear exchange reactions [10], in which a proton emits a virtual charged pion only when it "knows" there is a neutron to absorb it in the future. In this view, the rate of spontaneous emission of an excited atom floating in deep space is proportional to the density of potential absorbers in the infinite horizon. In this view, the rate of spontaneous emission can be changed – exactly as it is changed in more traditional CQED, which actually models the modes available within a cavity and does not really use what is assumed about the zero point field. In effect, the zero point field goes away in the predictions made for emission into a cavity. Practical CQED calculations generally analyze a two-level atom and the photon(s) in a cavity as a single system, subject to dissipation from the cavity following traditional master equation models of dissipation [13,15]. Dutra [14] points to an interesting theoretical literature on "open mode" modeling, which has great promise, but does not yet change the empirical picture here for optical quantum computing.

Many have suggested that the noise suppression power of CQED might be useful in quantum computing [3]. However, the traditional models [15] still clearly imply a time-forwards dissipation, following the good old fashioned forwards arrow of time, and the decoherence problem remains serious [3].

## 3.2. A Radical Approach

Is it really necessary that the arrow of time always go forwards, inescapably, in all parts of all engineering systems?

Indeed, if the underlying dynamic laws (such as a Schrodinger equation perhaps) which actually govern the universe are symmetric with respect to time, how could a time-forwards arrow of time exist at all? Are the emergent properties (such as the entropy function, a nonlocal function) of our universe really unrelated to the specific dynamics which give rise to them?

In [10], I propose that the usual arrow of time in everyday experience is an artifact based on *boundary conditions*. In all of our normal life and past experiments, we rely on properties of the boundary of far past time, a boundary property which presents itself to us as positive-time free energy. Yet logically, the underlying time-symmetric dynamics of the universe would clearly allow for the mirror image of this phenomenon, negative-time free energy. If we could find a source of negative-time free energy, we could reverse dissipation and noise locally, and enable an extension of QIS with a very powerful new degree of freedom. But where could we possibly find a source of negative time free energy?

## 3.3. A Possible Source of Negative Time Free Energy

Is there any way we could find a source of free energy which is not some kind of fossil energy from the far past? Price has suggested [16] that we try to look out in the universe, at least to try to see backwards light, perhaps from the "big crunch" at the other end of our universe; however, it seems unlikely that reverse-time solar energy would amount to much. Despite the obvious questions and uncertainties, the more promising approach would seem to be to try to extract energy somehow from heat, which looks the same in forwards or backwards time. (Note that ideas for exploiting zero point energy are a special case of this, as zero point energy – if it exists – would also be symmetric and disordered like heat.)

In 2002, a proposal for energy extraction from ambient heat arrived at NSF, from a well-respected empirical researcher in electronic devices. The panel rejected it, for obvious reasons, but he asked me: "Can you really prove mathematically that such things are impossible, in an interconnected universe governed by fields? I ask that you look into it and rely on what you yourself can prove, not on approximations or generalities they taught you in school."

In response, I reviewed what is actually known about the Second Law of Thermodynamics and about entropy in the general mathematical case [17]. I will not repeat all the details here – but in essence, a number of noted sources and logic all point towards a big loophole here, and a serious possibility of exploiting field effects to do things which previously seemed impossible. A number of serious experimenters hoped to prove this in replicable experiments of varying kinds, which really should have worked, according to standard quantum mechanics (KQED or FQED).

Yet a few years later, one of those lead experimenters was deeply puzzled as to why something which "had to work" did not actually work. "We have well-validated simulation models, based on standard QED, which clearly predict that it will work, but just for these systems the models break down and nothing works. You say you believe in a different version of quantum mechanics. Could you please look at this from the viewpoint of your new theory?"

So let us suppose some sort of photonic device, which converts ambient IR ("heat vibrations") into something more concentrated, and then connects that to a wire. The device could be like one of the many interesting systems studied at the Langevin Institute in France, or one of those described by Kong, Kim and Trew, or a spiral nanorectenna system, or a photonic crystal; there are many possibilities on a solid footing. What happens when it gets to the wire?

Normally, when we study electronic circuits, we assume that the external world provides a time-varying source of voltage at both terminals. Predicting the circuits is basically a glorified, enhanced calculation of electron transport [18,19], working out what happens when voltages at the terminals lead to certain probability distributions for the state of electrons in those terminals. (Unfortunately, these are called 'reservoirs' – quite different from the reservoirs of quantum optics [15].). The electron transport is basically an extension of the well-known Boltzmann transport equation [18], assuming a certain density of states $f(k)$ on the left for electrons with wave number k propagating from left to right, with positive k indicating motion from left to right.

In my theory, the terminal connecting the photonic system to the wire provides a high density *both* of electrons $f(k)$ and of electrons $f(-k)$ at the same terminal. In effect, it provides a negative voltage propagating forward in time, and a precisely equal positive voltage propagating backwards in time. In circuits designed to work with nonzero voltage, nothing happens. [9].

However, if we extend the circuit modeling to allow for such sources, we can still derive benefits from them. In particular, it is possible to develop a relatively simple quantum separator circuit, to separate streams of positive time free energy and negative time free energy. At the time when I figured out how to do this [20], I did not yet have a proper modeling framework – but I felt it was important to get the circuit concept documented and available. The new modeling framework in [7], [9] and [12] is more complex, but it seems clear that the circuit design should still work.

One of the difficulties in this research is that many experimental groups would prefer to work totally in photonics, without the connection to a wire. However, quantum separation requires fast switching, which is easier to do in today's state of the art electronics than in photonics. Even before a full-fledged quantum separator is constructed, use of one fast switch and a fast well-synchronized detector a meter downstream on the wire can test for the presence of the combination of forward-propagating modes and backward propagating modes predicted by time-symmetric physics, in the case where the photonic part achieves a high density of states.

The key issue now is how to find the photonic component which provides maximum local density of states forwards to the wire; that is a major research challenge, but now that we know what figure of merit to look for in the photonic component, one may hope that faster progress will be possible. This is a classic case of a high-risk high-potential opportunity, with many, many extensions possible if it can be made to work.